# Social value and information quality in online health information search

**Tahir Hameed**
SolBridge International School of Business
Woosong Educational Foundation
Daejeon, South Korea
Email: tahir@solbridge.ac.kr

**Bobby Swar**
SolBridge International School of Business
Woosong Educational Foundation
Daejeon, South Korea
Email: bswar@solbridge.ac.kr

## Abstract

This research-in-progress paper extends and validates a model of value-driven online health information (OHI) search and reuse in personal and online sharing contexts. Perceived value is an important factor behind users' decisions concerning search, consumption and reuse of products and services. The role of utilitarian, epistemic, and hedonic value of information is well-recognized in user satisfaction and subsequent intention to repeat online search. However, research till date show little support for social value, which is a critical driver of user satisfaction and decisions about continuance and information sharing. Therefore, a value-based model of OHI search is extended with detailed information quality and information value measures. Initial results from the survey data collected from 193 college students from 22 countries studying in South Korea demonstrated two novel results. At first, unlike existing studies, good support was found for the effect of perceived social value on user satisfaction. However, the significance of social value could only be appreciated by using comprehensive constructs of information quality. Secondly, it was noted when satisfied users make their decisions to repeat OHI search or information sharing on internet and digital social networks, they consider representation and context quality of health information more than intrinsic and access quality. Therefore, developers and organizations interested in traction of their internet and social media based healthcare information systems (HIS) should focus on enhancing the information quality of their websites and systems accordingly.

## 1  Introduction

Online health information (OHI) search is taking a new social life of its own in the internet (Fox 2011). It's not about searching and using information for self-consumption anymore, but also about searching, using and sharing it together with others. Health caregivers and organizations are also anticipating far-reaching consequences of social networking and media on how healthcare is planned, organized and delivered (Hesse et al. 2005; Lee 2012; Pradhan et al. 2013). Participatory medicine based on social media (Gallant et al. 2011), patients' peer- to peer experiences in digital healthcare communities (Ba et al. 2013) are just few examples. (Ziebland et al. 2012) notes, "The act of participating in the creation of health information also influences patients' experiences and has implications for our understanding of their role in their own health care management and information."

However, there are two major issues in adoption and continued use of OHI search and their sources especially those integrating social media. First, due to exploding volume and sources of online healthcare information, users find it hard to locate and use credible and valuable information which could effectively meet their search needs (Adams 2010; Eastin 2001; Zeng et al. 2004) . Useless or misleading information, which does not carry enough value or causes damages and unpleasant experiences (Kortum et al. 2008; White et al. 2009), could negatively affect user satisfaction levels and user intention for repeating OHI search. Alternatively, valuable information and pleasant OHI search experiences can lead users to come back searching for similar or different information. Secondly, if the users are satisfied, they would recommend and/or share such information with others via websites, blogs, wikis, or social networks (Fox 2011; Lee 2012; Pradhan et al. 2013) but only if they have enough motivation and incentives to do so.

Retailing literature introduced the role of perceived value in consumer behaviour about searching, consumption and reusing products and services (Sweeney et al. 2001; Zeithaml 1988). That inspired information systems (IS) researchers to explore the role of information value and if it affects the current and future information behaviour of online users, for instance in tourism, hospital and mobile entertainment services (Cai et al. 2004; Chahal et al. 2012; Petrick 2004; Pihlström et al. 2008).

Utilitarian value, epistemic (knowledge) value, and emotional value are known constructs of information value, which affect user satisfaction, and therefore should lead to repeated OHI search and online information sharing. However, social value is a construct of information value which has not received as much attention in the wake of digital social media technologies and their proliferation as it should have. (Goetzinger et al. 2007), which is the only study discussing social value in OHI search to the best of our knowledge, did not find a positive relationship between social value and user satisfaction therefore not providing any implications for OHI search in the online sharing contexts. That is very counter-intuitive in the context of new social life of health information use (Fox 2011) .

Therefore, this research raises and addresses two key concerns; first, a confirmation of whether or not social value of information is related to user satisfaction which could lead to the intention to repeat OHI search and/or sharing); second and more important if and how information quality has a role in defining social value of OHI in online sharing contexts.

To answer these questions, we extend and test the value-driven conceptual model by (Goetzinger et al. 2007). We employed relatively comprehensive constructs of information quality defined in the IS literature to confirm this relationship. Survey data from 193 college-going from 22 countries currently residing in South Korea for studies was collected. PLS-SEM Structural equation modelling software was used to test and validate the proposed model. The results not only show good support for social value as an integral construct of information value but they also highlight which aspects of information quality are relatively important for social value. Moreover, social value is positively related to user satisfaction and therefore should affect intentions for reuse and intentions for sharing of OHI search.

Rest of the paper is organized as follows. Section 2 introduces previous literature and theoretical background before setting up extended research model and testable hypotheses. Section 3 describes the instrument, data collection, results and analysis while section 4 provides brief conclusions, discussion and plans for future research.

## 2 THEORETICAL BACKGROUND AND RESEARCH MODEL

### 2.1 Online health information search behaviour

Information search was traditionally focused on efficiency and accuracy of information retrieval (IR) systems and their characteristics. However, information behaviour i.e. searching, using, reusing and sharing information is more pertinent for the internet era. Lately, information behaviour has been viewed from behavioural, cognitive, and staged-process viewpoints, for example see (Lambert et al. 2007; Wilson 1999). The decisions involving searching, evaluating, using and then reusing or sharing information are based on several factors, such as needs, individual's abilities, system characteristics, and risks causing variance in information behaviour (Dervin 1983; Ellis 1989; Wilson et al. 1996).

The use of online healthcare websites and applications is rapidly increasing. Everyday millions of internet searches are performed for health related information about general well-being, sexuality, nutrition, symptoms, diagnosis and treatments (Renahy et al. 2008). Pew research centre's American life and internet project (Fox et al. 2015) and their OHI search surveys, and other studies (Wartella et al. 2015) explain major predictors of OHI search include age, gender, internet literacy, specific health needs, belonging to online groups, caregiving, searcher's capabilities and social differences. (Escoffery et al. 2005; Morahan-Martin 2004; Rice 2006).

OHI search is changing the dynamics of healthcare delivery in several ways. For example, (Hesse et al. 2005) noted "Most physicians are already experiencing the effects of patients showing up to their offices armed with printouts from the World Wide Web and requesting certain procedures, tests, or medications." Recently. (Xiao et al. 2014) used 2003 HINTS data to demonstrate that access to online systems, trust levels, and health state of the users are related to frequency, diversity of usage, and preference of internet as medium when seeking OHI. Informed users could be more receptive to health information and treatments, however, there could also be cases of dissatisfaction or conflicts. Till date, the research shows individuals trust health professionals more than internet (Wartella et al. 2015).

Therefore, understanding online health information (OHI) search behaviour has critical implications for patients and caregivers in getting, using and sharing the right information, for professional care providers in guiding them to better information sources, and for companies developing healthcare delivery systems to improve the embedded information quality and value (Eysenbach et al. 2002).

A combined search about OHI and information systems (IS) on Google Scholar provided 209,000 publications. Generally, in IS literature, OHI search behaviour is approached via cross-cutting theories of IS implementation and adoption, media and communications studies, and healthcare psychology. For example, (Yoo et al. 2008) used theory of planned behaviour (Ajzen 1985; Ajzen 1991) and uses and gratification theory (Rayburn et al. 1984; Ruggiero 2000) together to study behavioural intentions of middle-aged women to information seeking on the internet. Similarly, (Yun et al. 2010), combined two well-known models i.e. technology acceptance model (TAM) (Davis 1989; Venkatesh et al. 2003) and health-belief model (Becker 1974; Janz et al. 1984) to study information-seeking behaviour of Korean adults about diseases on the internet.

Although most IS studies focus on the system characteristic as well as user context before or during the OHI search, they generally don't have constructs which could take into account all the factors related to adoption of choice objects; for a treatment see (Kim et al. 2007). Additionally, user satisfaction, gratification with specific media, and intention to repeat are based on assumptions that users are competent in evaluating the benefits they would get in the future. That might not be generally true about all, therefore necessitating better future-oriented constructs when repetition of earlier (buying) behaviour is expected. Moreover, there is a need to understand the post-usage decisions when user is more informed about marginal incentives and efforts in personal and sharing contexts. Value-driven models provide an alternative and better explanation in such cases (Chen et al. 2003; Kim et al. 2007).

## 2.2 Value-driven health information search and reuse

Perceived value is a recognized construct in marketing literature which leads to customer loyalty and reuse (Sánchez-Fernández et al. 2007; Zeithaml 1988). Customers' perceived value may vary for the product because there are elements of thinking and feeling than utility only, also including hedonic value (Batra et al. 1991; Sweeney et al. 2001). Moreover, perceived value is not limited to perceptions before or during the use only but also include a futuristic aspect. Therefore, linking the two notions with OHI search services, perceived information value can be defined as user perceptions of the tangible and intangible value from searching, using, repeating search and sharing online information in public domains, relative to one's contribution.

Perceived information value has important role in online information sharing contexts because users should involve in both consumption as well as production of information. It has several dimensions e.g. contextual value, conditional value, epistemic value, emotional value, social value, convenience value, and monetary value (Pihlström et al. 2008).

(Goetzinger et al. 2007) proposed a model of value-driven OHI search in which user satisfaction is driven by information value. They suggested information value, composed of epistemic value, utilitarian value and social value, is driven by high levels of 'information quality'. When searched information is high in relevance and clarity it leads to higher perceived information value. However, the two measures of information quality they used i.e. 'relevance' and 'clarity' are too broad to fully uncover social value. Therefore, it's not surprising they did not find a significant relation between social value and user satisfaction, hence no or little contribute of social value towards reuse.

## 2.3 Revised model with perceived social value of information in online sharing

Informaftion sharing was traditionally researched in formal or organizational knowledge management contexts, for example (Kimmerle et al. 2010; So et al. 2005). However, it is recognized that individuals learn also by seeking and sharing news, experiences, opinions, and expert advice on the internet via communities of practice, shared interest groups, expert blogs and wikis (Ba et al. 2013; Greene et al. 2011). (Wilson 1999) used the terms "Information Exchange" for information seeking from other people and "Information Transfer" for sharing the sought information with other people, but this paper generally considers both as information sharing. While there are plenty of studies on perceived value (Sánchez-Fernández et al. 2007), the conditions under which one decides to contribute efforts in information sharing on internet and social media still need investigation, particularly in online information sharing (Kimmerle et al. 2011).

Among other constructs of perceived information value, social value is particularly important for online information sharing because it is related to group behaviour in public domain where the assessment of perceived benefits and contributions is somewhat difficult due to information quality, unknown sources and variances in the feedback from others (Flanagin et al. 2014; Kimmerle et al. 2011). Additionally, individuals could be pro-social or pro-self who might be sharing information to gain social recognition or simply be altruistic, or even trying to improve the information quality for later use by self or others (Kimmerle et al. 2011).

Given extensive work on information quality measures for different contexts, for example (Lee et al. 2002), and the fact that research including (Goetzinger et al. 2007) did not find a significant relation between social value and user satisfaction, this paper argues that detailed information quality measures should be used to detect the perceived social value of information particularly in shared contexts, its relationship with user satisfaction and ensuing intentions to repeat and share OHI search.

Therefore, we attempt to revisit Goetzinger et al.'s (2007) model. It is based on the same premise that high level of *perceived information quality* is positively related to *perceived information value* which in turn is positively related to *user satisfaction* (US). Obviously, a higher level of US positively impacts behavioural intention to reuse. However, this paper limits itself leaving behavioural intention to reuse and sharing on the internet to a later version. Moreover, *perceived information value* will not be defined till late in this section, therefore for early reference, the model keeps the three constructs by Goetzinger et al. (2007), Epistemic Value (EV), Utilitarian Value (UV) and Social Value (SV).

Initially, we turn to the constructs of *perceived information quality* and their relationship with the *perceived information value* constructs. Although some studies provide specific measures of website quality or quality of information on the web (Aladwani et al. 2002; Wilson 2002), (Lee et al. 2002) developed a comprehensive set of information quality constructs, which we find more relevant for two reasons. First, they developed their constructs with a systematic approach covering a very detailed list of metrics. Second and more important, they merged academics' view of information quality with practitioners' view. In our opinion, such a view is much more suitable to consumer-oriented notion of perceived value while keeping the rigour for academic research.

They proposed four major constructs of *perceived information quality* which are incorporated in this paper. At first, Intrinsic quality (IQ) refers to the fact that information has basic quality of its own measured by *accuracy, reliability and correctness*. It is true for OHI that users frequently feel their inability to evaluate the correctness and reliability of information (Eysenbach et al. 2002). Therefore, healthcare websites use regulators' and accreditation marks in addition to citations in an attempt to improve their IQ by adding *reputation and good sources* (Wilson 2002). Consistency of information across different online sources also increases user confidence in the correctness of information, which could be measured as *similarity with other sources*. All the above measures therefore contribute to IQ and it could be hypothesized (H1) that correct, accurate and reliable information positively impacts user's knowledge, fulfilment of his OHI search objectives and his willingness to receive additional social benefits with marginal efforts. It is broken in three separate hypothesis.

*H1a: Intrinsic Quality (of health information) is positively related to Perceived Epistemic Value*

*H1b: Intrinsic Quality (of health information) is positively related to Perceived Utilitarian Value*

*H1c: Intrinsic Quality (of health information) is positively related to Perceived Social Value*

Second, Contextual Quality (CQ) measures aspects important to the performance of current task i.e. *relevance, completeness, appropriateness* and *timeliness*. They directly serve user's needs (Lee et al. 2002) and are expected to affect all three types of values positively, although stronger correlations could be expected for UV and SV than EV. However, healthcare websites are dynamic and it's hard the information providers to maintain the same levels of relevance and appropriateness (Flanagin et al. 2014). Therefore, some websites frequently cite the dates to inform users if the webpage was *updated* *recently*. Recently updated information is more relevant and timely, hence higher in CQ. Therefore, frequent updates would affect EV, UV and SV similarly in a positive way. Summing the ideas following three hypothesis are made about CQ's relations with information value:

*H2a: Contextual Quality (of health information) is positively related to Perceived Epistemic Value*

*H2b: Contextual Quality (of health information) is positively related to Perceived Utilitarian Value*

*H2c: Contextual Quality (of health information) is positively related to Perceived Social Value*

Third, Representational Quality (RQ) is about selection of media and formatting of the content for effectively conveying the information. Traditional studies place heavy emphasis on user interface of the websites and information formats. The metrics for RQ which overlap between academics' and practitioners' views include *interpretability, compactness, conciseness, ease of manipulation, and consistency of the format* across different webpages and artefacts (Kimmerle et al. 2009; Lee et al. 2002). If any of them measures low it leads to low RQ which in turn is detrimental to comprehension and usefulness of the information. Therefore, low RQ negatively affects learning and satisfaction of OHI search objectives. Users thank that sharing low RQ information may actually damage their social acceptance therefore lowering its SV. So, RQ is positively related to three types of information value:

*H3a: Representational Quality (of health information) is positively related to Perceived Epistemic Value*

*H3b: Representational Quality (of health information) is positively related to Perceived Utilitarian Value*

*H3c: Representational Quality (of health information) is positively related to Perceived Social Value*

Fourth and the final type of information quality is Access Quality (AQ). It measures *ease of access* to information, *provisions which enhance obtainability* of information, *speed* of OHI search, *compatibility or manipulability* of information formats. Use of multiple technology standards and media formats is therefore critical. Whereas high AQ solicits positive impact on UV and SV, the manipulability also brings some role for EV because synthesis and comparison of information from various sources has educational and cognitive aspects. Therefore, three hypothesis in this context are:

*H4a: Accessibility Quality (of health information) is positively related to Perceived Epistemic Value*

*H4b: Accessibility Quality (of health information) is positively related to Perceived Utilitarian Value*

*H4c: Accessibility Quality (of health information) is positively related to Perceived Social Value*

Next, we discuss three constructs for *perceived information value* mentioned above, Epistemic Value (EV), Utilitarian Value (UV) and Social Value (SV). For the first, when the user feels attainment of knowledge about health issues from OHI search results, which he/she could use again later, the type of value received is educational in nature, called epistemic value. Information high in EV affects US, even if the need of the user was served or not by that information, because the user feels more informed about specific health issues. Three key aspects to measure EV are *quality*, *contents* and the additional *guidance provided* by the information which increases user's knowledge and therefore user satisfaction, hence the hypothesis H5:

*H5: Perceived Epistemic Value (of health information) is positively related to User Satisfaction*

The second type of value, Utilitarian Value (UV) represents direct benefits expected by a user from the health information. For instance, if one could successfully find information about their particular symptoms and corresponding diagnosis, one would feel the purpose of search was achieved, hence to higher perception of UV. It could become even higher if the physician later confirms the tentative diagnosis from OHI search. Contrarily, if the user could not find the specific information successfully, or one has to go to several websites (or online information sources) then the UV for the OHI search would be lower. Therefore, the hypothesis H6 about UV's relation with US follow:

H6: *Perceived Utilitarian Value (of health information) is positively related to User Satisfaction*

Lastly, the Social Value (SV) of information represents the social benefits a user could get by sharing and interacting about the searched information on internet and social media networks e.g. social acceptability and recognition or even empathy and altruism in some cases. For instance, if a user perceives his/her contribution could enhance the information quality for others, or if one receives information of high quality because of other's contributions, one may feel highly satisfied and therefore choose to codify and share information in the public domain (Kimmerle et al. 2011; Moorhead et al. 2013). In that case, a user can not only access the same information again for self but may receive additional feedback and guidance. Therefore, reuse in the shared context could enhance overall information quality and SV itself (Kimmerle et al. 2011) which in turn enhances US. That leads us to H7.:

*H7: Perceived Social Value (of health information) is positively related to User Satisfaction*

The final construct is User Satisfaction (US) which represents perception of the extent to which a system met a user's requirements (Ives et al. 1983). In the OHI search context, it would be overall perception of how well the searched information and its sources me the user requirements. It begins with choosing to search the information online, and resulting exactness, usefulness, relevance, appropriateness and applicability of the information to OHI objectives (Goetzinger et al. 2007; Lee et al. 2002).

Summing up the above discussion, a revised value-based model of OHI search is presented in Figure 1.

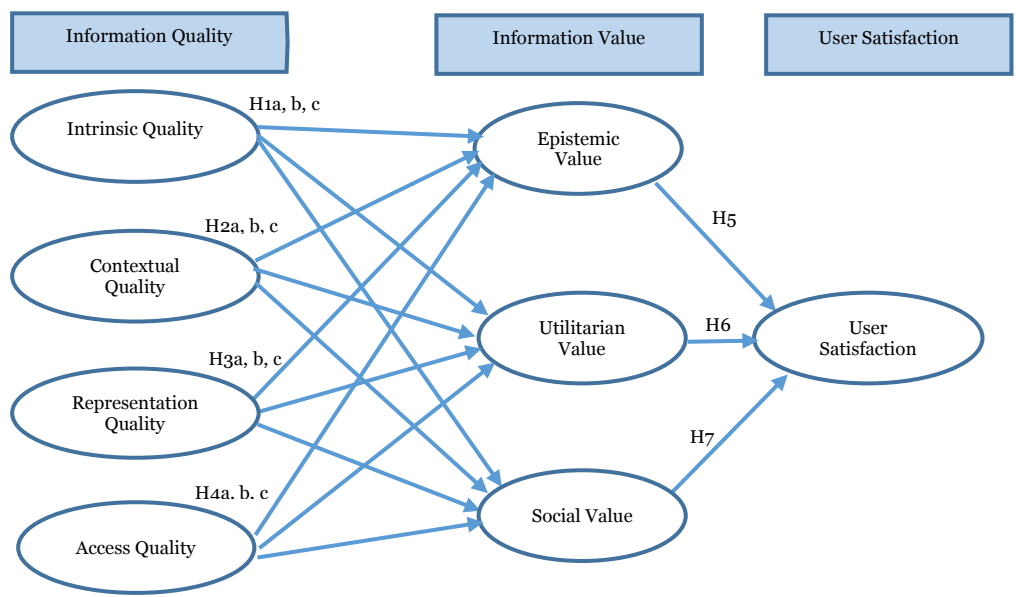

*Figure 1: A comprehensive value-driven model for OHI search for online sharing contexts*

## 2.4 Data collection and processing

A survey was developed based on 8 constructs in the model, each measured by several items (Babin et al. 1994; Ives et al. 1983; Lee et al. 2002; Sweeney et al. 2001; Venkatesh et al. 2003); See Appendix 1.

The survey was circulated to around 250 college-going youth from twenty-two different countries, all residing in South Korea for studies. Youth was targeted due to frequency of internet use, internet competency, social-media savviness, and concerns about health. Roughly 206 responses were received of which 193 were found usable. Structural Equation Modelling (PLS-SEM) was employed for data analysis. It requires at least 10 data points per path for a sample size significant for model testing (Hair et al. 2011). With 15 paths in above, 193 is an adequate sample size against the required 150. The results of the PLS-SEM are discussed in next section. Further data is under collection from users other than students and in other countries to add diversity to sample and predictive power in future research.

## 3　RESULTS

### 3.1　Assessment of the measurement model

Table 1 shows the assessment of the measurement model. Internal consistency reliability is examined by using composite reliability and Cronbach's alpha. The constructs in the proposed model are all above the recommended threshold value of 0.7 for both composite reliability and Cronbach's alpha, demonstrating high level of internal consistency reliability among all the latent variables (Hair et al. 2011). In the proposed model composite reliability values ranges from 0.865 (IQ) to 0.897 (US) and Cronbach's alpha value ranges from 0.771 (IQ) to 0.862 (US).

Convergent validity is assessed by evaluating Average Variance Extracted (AVE) from the measures. Table 1 shows that AVE values for measures ranges from 0.588 (AQ) to 0.719 (UV). The value for each latent variable's AVE are above the threshold value of 0.5 confirming the criteria of convergent validity.

Discriminant validity is assessed by examining the square root of AVE. Table 2 shows the square root of AVE of each construct is greater than the correlations between it and all other constructs. Also, all the constructs are found to have stronger correlation with their own measures than to those of others. This represents the proper assessment of discriminant validity.

| Variables | Cronbach's Alpha | AVE | Composite Reliability |
|---|---|---|---|
| Access Quality | 0.826 | 0.588 | 0.877 |
| Contextual Quality | 0.830 | 0.662 | 0.887 |
| Epistemic Value | 0.813 | 0.641 | 0.877 |
| Intrinsic Quality | 0.771 | 0.683 | 0.865 |
| Representational Quality | 0.808 | 0.638 | 0.875 |
| Social Value | 0.853 | 0.629 | 0.894 |
| User Satisfaction | 0.862 | 0.592 | 0.897 |
| Utilitarian Value | 0.805 | 0.719 | 0.885 |

*Table 1. Assessment of the Measurement Model*

|  | AQ | CQ | EV | IQ | RQ | SV | US | UV |
|---|---|---|---|---|---|---|---|---|
| AQ- Access Quality | **0.767** | | | | | | | |
| CQ- Contextual Quality | 0.601 | **0.814** | | | | | | |
| EQ- Epistemic Value | 0.527 | 0.584 | **0.801** | | | | | |
| IQ- Intrinsic Quality | 0.505 | 0.635 | 0.449 | **0.826** | | | | |
| RQ- Representational Quality | 0.572 | 0.462 | 0.455 | 0.370 | **0.799** | | | |
| SV- Social Value | 0.389 | 0.453 | 0.433 | 0.363 | 0.383 | **0.793** | | |
| US- User Satisfaction | 0.628 | 0.706 | 0.613 | 0.564 | 0.490 | 0.603 | **0.770** | |
| UV- Utilitarian Value | 0.581 | 0.530 | 0.485 | 0.461 | 0.491 | 0.330 | 0.596 | **0.848** |

*Table 2. Fornell-Lacker test of discriminant validity. Note: The diagonal elements (in bold) represent the square root of the AVE.*

## 3.2 Testing the model

Figure 2 depicts the results obtained from PLS-SEM analysis. The coefficient of determination, $R^2$ is 0.600 for US which means the structural model explained 60% of the variation in user satisfaction.

As shown in Figure 2 the relationship between IQ and UV (beta=0.123, $p<0.1$) is statistically significant supporting H1b. Contrary to the proposal, IQ's relationship with EV (beta=0.073) and SV (beta=0.086) are found insignificant hence H1a and H1c are not accepted.

CQ's relations are found to be statistically significant with EV (beta 0.351, $p<0.01$), UV (beta=0.184, $p<0.05$), and SV (beta=0.265, $p<0.01$), supporting H2a, H2b and H2c.

RQ is also found to have significant relationships with all types of values, EV (beta=0.158, p<0.05), UV (beta=0.188, p<0.05) and SV (beta=0.181, p<0.05), supporting H3a, H3b, and H3c respectively.

AQ is found to be significantly related with EV (beta=0.189, p<0.05) and UV (beta=0.301, p<0.01), supporting H4a and H4b. However, contrary to the assumptions, AQ has insignificant relationship with SV (beta=0.082).

Finally, EV (beta=0.293, p<0.01), UV (beta=0.334, p<0.01) and SV (beta=0.366, p<0.01) are all found to have statistically significant relationship with user satisfaction, supporting H5, H6 and H7.

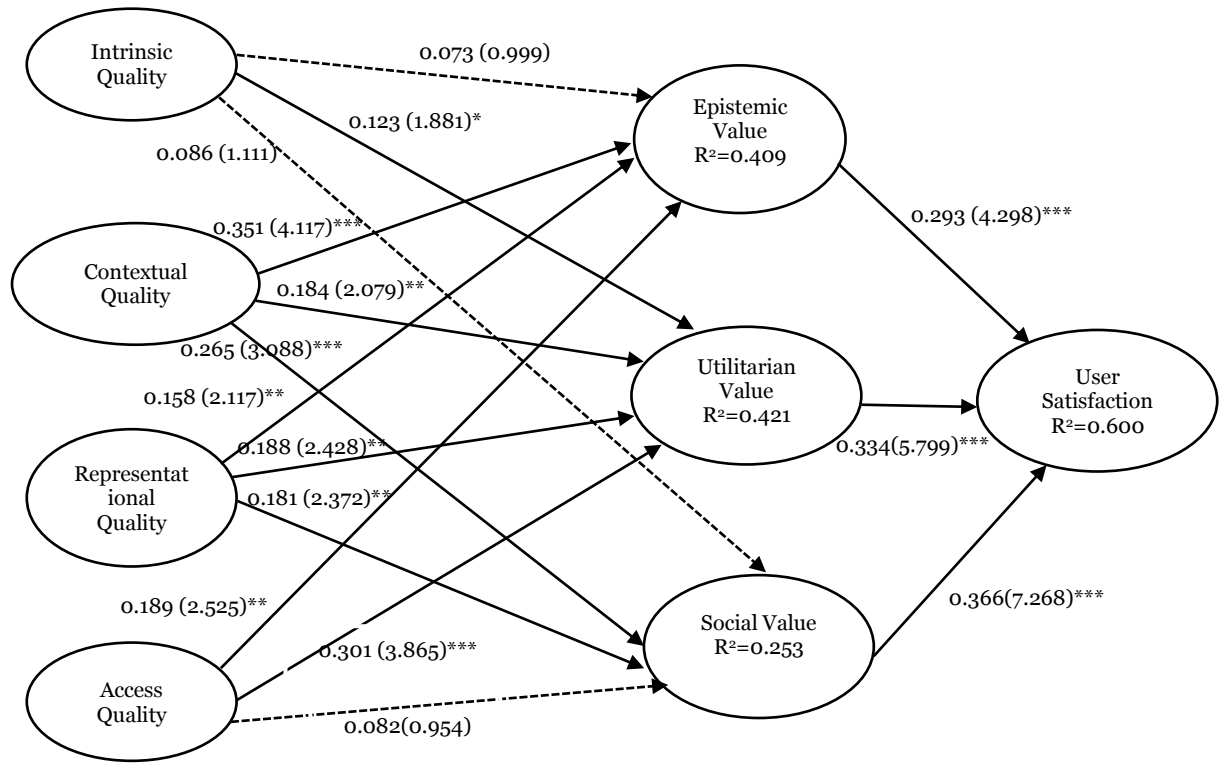

*Figure 2. Results of structural model with path coefficients (t-statistics in parentheses). Notes: \*p<0.1, \*\*p<0.05, \*\*\*p<0.01; Dashed lines represent non-significant relationships*

## 4   CONLCUSIONS, DISCUSSION AND FUTURE RESEARCH

Perceived information value affects user satisfaction. OHI seekers do care for the value they receive from the searched information whether educational, utilitarian or social. Generally, most aspects of information quality positively relate to user perceptions of information value.

It is evident information seekers see social value of their search results as a significant contributor to higher satisfaction levels, which implies they consider sharing the information or feedback about it on the internet and digital social media. Therefore, healthcare information systems developers and organizations should try improving those information quality aspects of their websites and social media systems which enhance social value for users. Satisfied users show a higher tendency for sharing OHI if it bears high contextual quality i.e. higher relevance, completeness, appropriateness and timeliness. They also feel more inclined to share if representational quality of information is high i.e. interpretability, compactness, conciseness, ease of manipulation, and consistency of the format.

Somewhat surprising, access quality in our results do not appear to push much sharing among users. The limitations of sample could be a cause behind insignificant relationships between CQ and SV. The respondents of current sample are internet competent and social media savvy users, who might consider their peers equally competent in internet search therefore did not feel any effects of access quality.

However, the insignificance of relationships between IQ (intrinsic or content quality) with two key constructs of information value i.e. SV and EV is even more surprising. It is hard to find out from the current data why information seekers would not value different measures of correctness and reliability of information for SV or EV. One reason could be low medical or health literacy of the respondents in current sample. Another reason could be perceived differences by the respondents about the OHI needs of individuals, which led them to believe correctness of the information might not be uniformly applicable to everyone. Yet another reason could be flawed survey instrument. Therefore, in all cases, future research should try revisiting the basis linking IQ with different constructs of information value. It is fairly possible that diversity of the sample could change these results. Therefore, authors plan to extend the data collection to respondents other than students, for example patients and medical students (as caregivers). Additionally, the data collection is also planned from other countries. New data would also enhance the validity and generalizability of the significant results.

Future research should provide an elaborate concept of social value for OHI and HIS which could lead to improvements in websites and digital social media systems design in different settings, for example user engagement with users, patients and possibly caregivers. It will be interesting to dig deeper how shared reuse interrelates with online communities of practice, online and off-line patient-caregiver interactions and patient-patient interactions.

A crude multi-group testing of the above data unearthed some effects of gender on the relationship between social value and user satisfaction. There could be a moderating role of gender on user satisfaction and hence shared use of OHI search and HIS. Similarly, there could be room for moderating roles of cultural differences, language competencies and healthcare knowledge competencies.

Social identity literature shows there could be strong relationships between perceived social value and social identity of the user. Users with strong ties and critical nodes in social networks might have differences in perceived social value from their counterparts in general. That might also be an avenue for research for the interested.

# 5  References

Adams, S. A. 2010. "Revisiting the online health information reliability debate in the wake of "web 2.0": an inter-disciplinary literature and website review," *International journal of medical informatics* (79:6), pp 391-400.

Ajzen, I. 1985. *From intentions to actions: A theory of planned behavior*, (Springer.

Ajzen, I. 1991. "The theory of planned behavior," *Organizational behavior and human decision processes* (50:2), pp 179-211.

Aladwani, A. M., and Palvia, P. C. 2002. "Developing and validating an instrument for measuring user-perceived web quality," *Information & management* (39:6), pp 467-476.

Ba, S., and Wang, L. 2013. "Digital health communities: The effect of their motivation mechanisms," *Decision Support Systems* (55:4), pp 941-947.

Babin, B. J., Darden, W. R., and Griffin, M. 1994. "Work and/or fun: measuring hedonic and utilitarian shopping value," *Journal of consumer research*), pp 644-656.

Batra, R., and Ahtola, O. T. 1991. "Measuring the hedonic and utilitarian sources of consumer attitudes," *Marketing letters* (2:2), pp 159-170.

Becker, M. H. 1974. *The health belief model and personal health behavior*, (Slack.

Cai, L. A., Feng, R., and Breiter, D. 2004. "Tourist purchase decision involvement and information preferences," *Journal of Vacation Marketing* (10:2), pp 138-148.

Chahal, H., and Kumari, N. 2012. "Consumer perceived value: The development of a multiple item scale in hospitals in the Indian context," *International Journal of Pharmaceutical and Healthcare Marketing* (6:2), pp 167-190.

Chen, Z., and Dubinsky, A. J. 2003. "A conceptual model of perceived customer value in e-commerce: A preliminary investigation," *Psychology & Marketing* (20:4), pp 323-347.

Davis, F. D. 1989. "Perceived usefulness, perceived ease of use, and user acceptance of information technology," *MIS quarterly*), pp 319-340.

Dervin, B. 1983. *An overview of sense-making research: Concepts, methods, and results to date*, (The Author.

Eastin, M. S. 2001. "Credibility assessments of online health information: The effects of source expertise and knowledge of content," *Journal of Computer-Mediated Communication* (6:4), pp 0-0.

Ellis, D. 1989. "A behavioural approach to information retrieval system design," *Journal of documentation* (45:3), pp 171-212.

Escoffery, C., Miner, K. R., Adame, D. D., Butler, S., McCormick, L., and Mendell, E. 2005. "Internet use for health information among college students," *Journal of American College Health* (53:4), pp 183-188.

Eysenbach, G., and Köhler, C. 2002. "How do consumers search for and appraise health information on the world wide web? Qualitative study using focus groups, usability tests, and in-depth interviews," *Bmj* (324:7337), pp 573-577.

Flanagin, A. J., Hocevar, K. P., and Samahito, S. N. 2014. "Connecting with the user-generated Web: how group identification impacts online information sharing and evaluation," *Information, Communication & Society* (17:6), pp 683-694.

Fox, S. 2011. *The social life of health information 2011*, (Pew Internet & American Life Project Washington, DC.

Fox, S., and Duggan, M. 2015. "Pew Internet and American Life Project," Pew Research Center Washington, DC.

Gallant, L. M., Irizarry, C., Boone, G., and Kreps, G. L. 2011. "Promoting participatory medicine with social media: new media applications on hospital websites that enhance health education and e-patients' voices," *Journal of participatory medicine* (3), p e49.

Goetzinger, L., Park, J., Jung Lee, Y., and Widdows, R. 2007. "Value-driven consumer e-health information search behavior," *International Journal of Pharmaceutical and Healthcare Marketing* (1:2), pp 128-142.

Greene, J. A., Choudhry, N. K., Kilabuk, E., and Shrank, W. H. 2011. "Online social networking by patients with diabetes: a qualitative evaluation of communication with Facebook," *Journal of general internal medicine* (26:3), pp 287-292.

Hair, J. F., Ringle, C. M., and Sarstedt, M. 2011. "PLS-SEM: Indeed a silver bullet," *Journal of Marketing Theory and Practice* (19:2), pp 139-152.

Hesse, B. W., Nelson, D. E., Kreps, G. L., Croyle, R. T., Arora, N. K., Rimer, B. K., and Viswanath, K. 2005. "Trust and sources of health information: the impact of the Internet and its implications for health care providers: findings from the first Health Information National Trends Survey," *Archives of internal medicine* (165:22), pp 2618-2624.

Ives, B., Olson, M. H., and Baroudi, J. J. 1983. "The measurement of user information satisfaction," *Communications of the ACM* (26:10), pp 785-793.

Janz, N. K., and Becker, M. H. 1984. "The health belief model: A decade later," *Health Education & Behavior* (11:1), pp 1-47.

Kim, H.-W., Chan, H. C., and Gupta, S. 2007. "Value-based adoption of mobile internet: an empirical investigation," *Decision Support Systems* (43:1), pp 111-126.

Kimmerle, J., and Cress, U. 2009. "Visualization of Group Members' Participation How Information-Presentation Formats Support Information Exchange," *Social Science Computer Review* (27:2), pp 243-261.

Kimmerle, J., Cress, U., and Held, C. 2010. "The interplay between individual and collective knowledge: technologies for organisational learning and knowledge building," *Knowledge Management Research & Practice* (8:1), pp 33-44.

Kimmerle, J., Wodzicki, K., Jarodzka, H., and Cress, U. 2011. "Value of information, behavioral guidelines, and social value orientation in an information-exchange dilemma," *Group Dynamics: Theory, Research, and Practice* (15:2), p 173.

Kortum, P., Edwards, C., and Richards-Kortum, R. 2008. "The impact of inaccurate Internet health information in a secondary school learning environment," *Journal of medical Internet research* (10:2).

Lambert, S. D., and Loiselle, C. G. 2007. "Health information—seeking behavior," *Qualitative health research* (17:8), pp 1006-1019.

Lee, R. 2012. "The rise of the e-patient, understanding social networks and online health information seeking," *Pew Internet Project, Pew Research Center*).

Lee, Y. W., Strong, D. M., Kahn, B. K., and Wang, R. Y. 2002. "AIMQ: a methodology for information quality assessment," *Information & management* (40:2), pp 133-146.

Moorhead, S. A., Hazlett, D. E., Harrison, L., Carroll, J. K., Irwin, A., and Hoving, C. 2013. "A new dimension of health care: systematic review of the uses, benefits, and limitations of social media for health communication," *Journal of medical Internet research* (15:4).

Morahan-Martin, J. M. 2004. "How internet users find, evaluate, and use online health information: a cross-cultural review," *CyberPsychology & Behavior* (7:5), pp 497-510.

Petrick, J. F. 2004. "First timers' and repeaters' perceived value," *Journal of Travel Research* (43:1), pp 29-38.

Pihlström, M., and Brush, G. J. 2008. "Comparing the perceived value of information and entertainment mobile services," *Psychology & Marketing* (25:8), pp 732-755.

Pradhan, S., Gay, V., and Nepal, S. 2013. "Social Networking and Dental Care: State of the Art and Analysis of the Impact on Dentists, Dental Practices and their Patients," *BLED 2013 Proceedings*).

Rayburn, J. D., and Palmgreen, P. 1984. "Merging uses and gratifications and expectancy-value theory," *Communication Research* (11:4), pp 537-562.

Renahy, E., Parizot, I., and Chauvin, P. 2008. "Health information seeking on the Internet: a double divide? Results from a representative survey in the Paris metropolitan area, France, 2005–2006," *BMC Public Health* (8:1), p 69.

Rice, R. E. 2006. "Influences, usage, and outcomes of Internet health information searching: multivariate results from the Pew surveys," *International journal of medical informatics* (75:1), pp 8-28.

Ruggiero, T. E. 2000. "Uses and gratifications theory in the 21st century," *Mass communication & society* (3:1), pp 3-37.

So, J. C., and Bolloju, N. 2005. "Explaining the intentions to share and reuse knowledge in the context of IT service operations," *Journal of Knowledge Management* (9:6), pp 30-41.

Sweeney, J. C., and Soutar, G. N. 2001. "Consumer perceived value: The development of a multiple item scale," *Journal of retailing* (77:2), pp 203-220.

Sánchez-Fernández, R., and Iniesta-Bonillo, M. Á. 2007. "The concept of perceived value: a systematic review of the research," *Marketing theory* (7:4), pp 427-451.

Venkatesh, V., Morris, M. G., Davis, G. B., and Davis, F. D. 2003. "User acceptance of information technology: Toward a unified view," *MIS quarterly*), pp 425-478.

Wartella, E., Rideout, V., Zupancic, H., Beaudoin-Ryan, L., and Lauricella, A. 2015. "Teens, Health, and Technology: A National Survey," Northwestern University.

White, R. W., and Horvitz, E. 2009. "Cyberchondria: studies of the escalation of medical concerns in web search," *ACM Transactions on Information Systems (TOIS)* (27:4), p 23.

Wilson, P. 2002. "How to find the good and avoid the bad or ugly: a short guide to tools for rating quality of health information on the internet," *BMJ: British Medical Journal* (324:7337), p 598.

Wilson, T. D. 1999. "Models in information behaviour research," *Journal of documentation* (55:3), pp 249-270.

Wilson, T. D., and Walsh, C. 1996. *Information behaviour: An inter-disciplinary perspective: A review of the literature*, (British Library Research and Innovation Centre London.

Xiao, N., Sharman, R., Rao, H. R., and Upadhyaya, S. 2014. "Factors influencing online health information search: An empirical analysis of a national cancer-related survey," *Decision Support Systems* (57), pp 417-427.

Yoo, E. Y., and Robbins, L. S. 2008. "Understanding middle-aged women's health information seeking on the web: A theoretical approach," *Journal of the American Society for Information Science and Technology* (59:4), pp 577-590.

Yun, E. K., and Park, H. 2010. "Consumers' disease information–seeking behaviour on the Internet in Korea," *Journal of clinical nursing* (19:19-20), pp 2860-2868.

Zeithaml, V. A. 1988. "Consumer perceptions of price, quality, and value: a means-end model and synthesis of evidence," *The Journal of marketing*), pp 2-22.

Zeng, Q. T., Kogan, S., Plovnick, R. M., Crowell, J., Lacroix, E.-M., and Greenes, R. A. 2004. "Positive attitudes and failed queries: an exploration of the conundrums of consumer health information retrieval," *International journal of medical informatics* (73:1), pp 45-55.

Ziebland, S., and Wyke, S. 2012. "Health and illness in a connected world: how might sharing experiences on the internet affect people's health?," *Milbank Quarterly* (90:2), pp 219-249.

**Appendix 1: Survey Questions (Likert Scale 1-Lowest, 5-Highest)**

| | | |
|---|---|---|
| **Information Quality** | **Intrinsic Quality (IQ):** The information I found and used | is correct. |
| | | is accurate. |
| | | is reliable. |
| | **Context Quality (CQ):** The information I found and used | is useful to my need. |
| | | is relevant to my need. |
| | | is appropriate for my need. |
| | | is applicable to my need. |
| | **Representation Quality (RQ):** The information I found and used | is easy to understand. |
| | | is formatted compactly. |
| | | is presented concisely. |
| | | is presented consistently in the same format. |
| | **Access Quality (AQ):** The information I found and used | is easily accessible. |
| | | is easily obtainable. |
| | | is quickly accessible when needed. |
| | | is easy to manipulate to meet my needs. |
| | | is easy to combine with other information. |
| **Information Value** | **Utilitarian Value (UV):** When I searched and used the information on the internet | I could get the exact information I wanted. |
| | | I found the information I was looking for. |
| | | I felt this information search was successful. |
| | **Epistemic Value (EV):** When I searched and used the information on the internet | The quality of information influences my knowledge about health. |
| | | The contents of information influences my knowledge about health. |
| | | The guidance received from information affects my knowledge about health. |
| | | I learn new things from the information. |

|  | **Social Value (SV):** When I searched, used and/or shared the information on the internet | would help me to feel acceptable. |
|---|---|---|
|  |  | would improve the way I am perceived. |
|  |  | would make a good impression of me. |
|  |  | would provide me with social approval. |
|  |  | would make me feel better that the feedback received from other people could help me. |
| **User Satisfaction** | **User Satisfaction (US):** When I searched and used the information on the internet | I think that I did the right thing when I searched online health information. |
|  |  | is exactly what was needed. |
|  |  | is useful to address my needs, concerns and objectives. |
|  |  | is relevant to address my needs, concerns and objectives. |
|  |  | is appropriate to address needs, concerns and objectives. |
|  |  | is applicable to address needs, concerns and objectives. |